\documentclass[a4paper,superscriptaddress,aps,prb,twocolumn,floatfix,citeautoscript]{revtex4-1}
\usepackage[utf8]{inputenc}
\usepackage{amsmath}
\usepackage{amssymb}
\usepackage{graphicx}
\usepackage{cleveref}
\usepackage{xcolor}
\usepackage{cancel}
\usepackage{booktabs}
\usepackage{ulem}

\usepackage{feynmp-auto}
\usepackage{lipsum}

\newcommand{\lcpq}{Laboratoire de Chimie et Physique Quantiques, Universit\'e de Toulouse, UPS, CNRS, and European Theoretical Spectroscopy Facility (ETSF), 118 route de Narbonne, F-31062 Toulouse, France}
\newcommand{\lpt}{Laboratoire de Physique Th\'eorique, CNRS, Universit\'e de Toulouse, UPS, and European Theoretical Spectroscopy Facility (ETSF), 118 route de Narbonne, F-31062 Toulouse, France}
\newcommand{\wisconsin}{Department of Materials Science and Engineering, University of Wisconsin-Madison, 53706, USA}
\newcommand{\ldeuxc}
{Laboratoire Charles Coulomb (L2C), Universit\'e de Montpellier, CNRS, F-34095, Montpellier, France}

\begin{document}
\title{Multichannel Dyson equations for even- and odd-order Green's functions: application to double excitations}

\author{Gabriele Riva}
\email{griva@irsamc.ups-tlse.fr}
\affiliation{\wisconsin}
\author{Théodore Fischer}
\affiliation{\ldeuxc}
\author{Stefano Paggi}
\affiliation{\lpt}
\author{J. Arjan Berger}
\email{arjan.berger@irsamc.ups-tlse.fr}
\affiliation{\lcpq}
\author{Pina Romaniello}
\email{pina.romaniello@irsamc.ups-tlse.fr}
\affiliation{\lpt}

\begin{abstract}
We extend the concept of the multichannel Dyson equation that we have recently derived to model photoemission spectra by coupling the one- and the three-body Green's functions, to higher-order Green's functions and to other spectroscopies. We show the general structure of the equations and how one can systematically approximate the corresponding multichannel self-energy. As a particular case, we focus on the coupling of the two-body and the four-body Green's functions in the electron-hole channel to describe neutral excitations. This formulation allows for the description of important many-body effects, such biexcitons, in a natural way. We illustrate our approach by applying it to a two-level model system, which, in a one-particle picture, exhibits single and double excitations. Our method can correctly describe both kinds of excitation, unlike standard approaches, and in good agreement with the exact results.
\end{abstract}

\maketitle
\section{Introduction}
Many-body perturbation theory based on Green’s functions \cite{Hed65,Bec16,martin_reining_ceperley_2016} plays a central role in the theoretical description of material properties thanks to the straightforward connection of the Green's functions to many properties of interest. For example, photoemission or optical absorption spectra are linked to the one-body Green's function (1-GF) and two-body  Green's function (2-GF), respectively.
Both the 1-GF and the 2-GF are calculated by solving the corresponding Dyson-like equation which links the non-interacting GF to the interacting one through an effective potential, which in general is frequency dependent. Thanks to this frequency dependence, additional spectral features that are not present at the non-interacting level, such as plasmon and double-plasmon excitations, or biexcitons, can be described in the many-body spectrum. However, in practice, there are several drawbacks: 1) standard frequency-dependent approximations do not yield very accurate results, in particular for satellites\cite{martin_reining_ceperley_2016,Blomberg72,Ary98,Kheifets_PRB03,Guz11,Dis15,Dis16,Dis21-2}, since they are not treated on equal footing with quasi-particles~\cite{riva_prl}; 2) a frequency-dependent approximation makes the solution of the Dyson equation cumbersome, in particular if it is solved self-consistently~\cite{Stan_2009,Rostgaard_2010,Car12,Car13, Kut10,Kut16,Kut17,Gru18,Yeh_2022}, and could lead to unphysical solutions~\cite{Lan12,Ber14,Sta15,Tar17,Loo18,Ver18,Ber21}; 3) a fully dynamical Dyson equation can be very complicated, as is the case, for example, for the dynamical Bethe-Salpeter equation of the 2-GF. ~\cite{Stri88,Rohlfing_2000,Rom09-2,San11,martin_reining_ceperley_2016,Authier_2020,Loos_2020,Bintrim_2022}

In this work, we build on a strategy that some of us have recently proposed to obtain photoemission spectra from the three-body Green's function.~\cite{Riv22,riva_prl,riva_prb} This strategy employs a multichannel Dyson equation (MCDE) which couples independent-particle one-body and three-body Green's functions through a multichannel self-energy. In our approach, quasiparticles and satellites are put on equal footing and, therefore, accurate results could be obtained from relatively simple approximations to the multichannel self-energy. For example, a static multichannel self-energy only containing the bare Coulomb interaction yields exact spectral functions of the symmetric Hubbard dimer~\cite{riva_prl}, contrary to standard methods such as $GW$ or the T-matrix approximation \cite{Rom09-1,Rom12}. 
This strategy follows a well-known idea: by starting from a large space, a static self-energy is able to capture the dynamical properties of a smaller space.~\cite{Sch78,Sch83,Nie84}

In this work, we generalize the MCDE to the coupling of more than two $n$-body Green's functions ($n$-GFs) and to other spectroscopies. As we will show in the following, in our approach $n$-GFs with $n$ even do not couple to those with $n$ odd. Therefore, multichannel Dyson equations are naturally split into two classes, those involving only $n$-GFs with $n$ even and those involving only $n$-GFs with $n$ odd.
As a particular case, we focus here on the MCDE that couples the electron-hole channel of the 2-GF and the 2-electron-2-hole channel of the 4-GF, which can be used to simulate absorption spectra and, in particular, spectra with features that have a double-excitation character. 
The standard approach to describe absorption spectra within many-body perturbation theory is the Bethe-Salpeter equation (BSE), which is a Dyson equation for the 2-GF.\cite{Stri88,Rohlfing_2000,Oni02,martin_reining_ceperley_2016} However the kernel of this equation is usually approximated to be static, which means that one cannot describe multi-excitation effects such those mentioned above. 
Our approach, instead, offers a natural framework to couple excitations.
In this context it is worth mentioning the cumulant approach, which seems promising both for the description of photoemission spectra and optical spectra \cite{Guz11,Cud20}. We finally mention the work of Torche and Bester which extends the BSE to the 2-electron-2-hole space  to target biexcitons.\cite{Torche_2021}

The paper is organized as follows.
In section \ref{sec:theory} we give the general structure of the MCDE for the coupling of even and odd $n$-GFs. We then focus on the 4-GF and we discuss its physical content, its MCDE and the corresponding multichannel self-energy. We perform a diagrammatic analysis of the MCDE and show how it can be solved in practice by rewriting the MCDE in terms of an eigenvalue equation with an effective Hamiltonian. In section \ref{sec:applications} we illustrate our approach using a simple Helium-like two-level model for which we can calculate the exact and MCDE neutral excitations. 
Finally, we draw our conclusions in section \ref{sec:conclusions}.
%
%
\section{Theory\label{sec:theory}}
\subsection{Multi-channel Dyson equations}
Let us consider an $N$-electron system in its initial ground state that upon a perturbation evolves to a final state with $M$ electrons. Let us first consider the case $M=N\pm 1$. It is well-known that the evolution of the system due to such a perturbation can be described by the 1-body Green's function (1-GF). In particular, the poles of the 1-GF correspond to the removal and addition energies of one electron. However, the evolution can also be described by the 
2-electron-1-hole ($2e$-$1h$) and 2-hole-1-electron ($2h$-$1e$) channels of the 3-GF since these channels correspond to the same $M=N\pm 1$ final state. 
Indeed, it can be shown that these channels of the 3-GF have the same poles as the 1-GF.~\cite{Riv22}
Instead, the amplitudes corresponding to the poles are, in general, different.
In general, the $ke$-$(k+1)h$ and $kh$-$(k+1)e$ channels of any $(2k+1)$-GF correspond to an $M=N\pm 1$ final state and they all have the same poles as the 1-GF.

Similarly, for the case $M=N$ it is well-known that the evolution of the system corresponding to such a perturbation can be described by the $1e$-$1h$ channel of the 2-GF. However, the $ke$-$kh$ channel of any $(2k)$-GF corresponds to an $M=N$ final state.
Moreover, they all have the same poles as the $1e$-$1h$ channel of the 2-GF,
e.g., the $2e$-$2h$ channel of the 4-GF (see section \ref{G4:Sec}).
Since channels of $n$-GFs corresponding to different final states describe different physical processes we can define a multichannel Dyson equation (MCDE) for each final state $M=N\pm s$.
In this work we will focus on the cases $s=0$ and $s=1$ but other MCDE can be derived for $s>1$.
In an MCDE several independent-particle (IP) $n$-GFs, e.g., noninteracting or Hartree-Fock (HF) $n$-GFs, are coupled through a multichannel self-energy. The starting point to derive an MCDE is the following Dyson equation
\begin{equation}\label{multiDyson:eq}
    G_n(\omega)=G^0_{n}(\omega)+G^0_{n}(\omega)\Sigma_n(\omega)G_n(\omega),
\end{equation}
where $G^0_{n}(\omega)$ is an IP $n$-GF and $\Sigma_n(\omega)$ is the $n$-body self-energy, which is defined by Eq.~(\ref{multiDyson:eq}).

It is convenient to project Eq.~(\ref{multiDyson:eq}) on the one-particle basis set $\{\phi_i\}$ that diagonalizes $G_n^0$. 
For an $M=N\pm 1$ final state $G_n^0$ then reads
\begin{align}
  G^0_{n}(\omega)=\begin{pmatrix}\label{Eqn:G0n+1}
        G^{0,1\text p}(\omega) & 0 & \cdots & 0\\
        0 & G^{0,3\text p}(\omega)& \cdots & 0  \\
        \vdots & \vdots & \ddots & 0\\
        0 & 0 & 0& G^{0,n\text p}(\omega)
    \end{pmatrix},
\end{align}
with
\begin{widetext}
\begin{align}
    G_{i_1;j_1}^{0, 1\text p}(\omega)&=
    \frac{\delta_{i_1j_1}}
    {\omega-\epsilon^0_{i_1}+i\eta\text{sign}(\epsilon^0_{i_1}-\mu)}\\
    G_{i_1\cdots i_n;j_1\cdots j_n}^{0, n\text p}(\omega)&=
    \frac{\delta_{i_1j_1}\cdots\delta_{i_nj_n}
    f_{i_1i_{(n+1)/2+1}} \cdots f_{i_1i_n}
    f_{i_2i_{(n+1)/2+1}} \cdots f_{i_{(n+1)/2 i_n}}}
    {\omega-\epsilon^0_{i_1}-\Delta\epsilon^0_{i_2i_{(n+1)/2+1}} \cdots-\Delta\epsilon^0_{i_{(n+1)/2}i_n}+i\eta\text{sign}(\epsilon^0_{i_1}-\mu)} \quad\quad (n\ge 3),
    \label{G033p:eq}
\end{align}
\end{widetext}
in which $\epsilon^0_i$ and $f_i$ are the energy and occupation number, respectively, corresponding to $\phi_i$, $\Delta\epsilon_{jk} = \epsilon^0_j - \epsilon^0_k$, $f_{kj} = f_k - f_j$, and $\mu$ is the chemical potential. 
In the following we will suppress the subscript of an $n$-body Green's function $G_n$ whenever the number of arguments or indices of the Green function are explicitly given, e.g., $G_{3\,ijl;mok} \to G_{ijl;mok}$.
Instead, for an $M=N$ final state $G_n^0$ is given by
\begin{align}
  G^0_{n}(\omega)=\begin{pmatrix}\label{Eqn:G0n}
        G^{0,2\text p}(\omega) & 0 & \cdots & 0\\
        0 & G^{0,4\text p}(\omega)& \cdots & 0  \\
        \vdots & \vdots & \ddots & 0\\
        0 & 0 & 0& G^{0,n\text p}(\omega)
    \end{pmatrix},
\end{align}
with
\begin{widetext}
\begin{align}
\label{G0n2p:eq}
    G_{i_1i_2;j_1j_2}^{0, 2\text p}(\omega)&=- i
   \frac{\delta_{i_1j_1}\delta_{i_2j_2}
  f_{i_2i_1}
    }{\omega-\Delta\epsilon^0_{i_1i_2}+i\eta\text{sign}f_{i_{2}i_1}}.
    \\
    G_{i_1\cdots i_n;j_1 \cdots j_n}^{0, n\text p}(\omega)&=- i
   \frac{\delta_{i_1j_1}\cdots\delta_{i_nj_n}
  f_{i_ni_1}\cdots f_{i_{n/2+1}i_1}f_{i_{n-1}i_2}\cdots f_{i_{n/2+1}i_{n/2}}  
    }{\omega-\Delta\epsilon^0_{i_1i_n}-\Delta\epsilon^0_{i_2i_{n-1}}\cdots-\Delta\epsilon^0_{i_{n/2}i_{n/2+1}}+i\eta\text{sign}f_{i_{n}i_1}} \quad\quad (n\ge 4).
    \label{G0nnp:eq}
\end{align}
\end{widetext}
We note that in Eq.~\eqref{G033p:eq} the restrictions $i_1>i_2>\cdots > i_{(n+1)/2}$ and $i_{(n+1)/2+1} > \cdots > i_n$ and similar restrictions on $j_1 \cdots j_n$ to avoid double counting are implicit~\cite{riva_prl}.
For the same reason, in Eq.~\eqref{G0nnp:eq} the restrictions $i_1>i_2>\cdots > i_{n/2}$ and $i_{n/2+1} > \cdots > i_n$ and similar restrictions on $j_1 \cdots j_n$ are implicit.
As a consequence, $G^0_{n}(\omega)$ in Eqs.~\eqref{Eqn:G0n+1} and \eqref{Eqn:G0n} is invertible~\cite{riva_prl}.
Moreover, equations \eqref{Eqn:G0n+1} and \eqref{Eqn:G0n} thus define the spaces in which the Dyson equation (\ref{multiDyson:eq}) has to be solved. 
With the definitions of $G^0_n$ given in Eqs.~\eqref{Eqn:G0n+1} and \eqref{Eqn:G0n}, the Dyson
equation in Eq.~\eqref{multiDyson:eq} becomes a multichannel Dyson equation, in which the multichannel self-energy is defined as
\begin{align}\label{selfmultiodd:eq}
\Sigma_n=\begin{pmatrix}
        \Sigma^{1\text{p}} & \Sigma^{ 1 \text p / 3\text p} &0  & \cdots &0\\
        \Sigma^{ 3 \text p / 1\text p} & \Sigma^{3\text{p}} & \Sigma^{ 3 \text p / 5\text p} & ...& 0\\
        0 &  \Sigma^{ 5 \text p / 3\text p} & \ddots & \cdots & 0\\
        \vdots & \vdots &\vdots &\Sigma^{(n-2)\text{p}} & \Sigma^{ (n-2) \text p / n\text p} \\
        0& 0 & 0 &\Sigma^{ n \text p / (n-2)\text p}&\Sigma^{n\text p}
    \end{pmatrix},
\end{align}
if $n$ is odd, and as
\begin{align}\label{selfmultieven:eq}
\Sigma_n=
\begin{pmatrix}
        \Sigma^{2\text{p}} & \Sigma^{ 2 \text p / 4\text p} &0  & \cdots &0 \\
        \Sigma^{ 4 \text p / 2\text p} & \Sigma^{4\text{p}} & \Sigma^{ 4 \text p / 6\text p} & ...& 0\\
        0 &  \Sigma^{ 6 \text p / 4\text p} & \ddots & \cdots & 0\\
        \vdots & \vdots &\vdots &\Sigma^{(n-2)\text{p}} & \Sigma^{ (n-2) \text p / n\text p} \\
        0& 0 & 0 &\Sigma^{ n \text p / (n-2)\text p}&\Sigma^{n\text p}
    \end{pmatrix},
\end{align}
if $n$ is even. 

In practice, when the number of electrons is large, we have to truncate the MCDE by retaining only the $n$-body Green's functions of lowest order. In general, an MCDE thus depends on the final state and the number of Green's functions that are coupled. To distinguish the various MCDE we will refer to each MCDE as the $(n,s)$-MCDE with $n$ the highest-order Green's function in the MCDE and $s=|M-N|$ with $N$ and $M$ the number of electrons of the initial and final state, respectively. 
For example, the MCDE that couples the 1-GF with the $2e$-$1h$ and $2h$-$1e$ channels of the 3-GF will be denoted $(3,1)$-MCDE
and the MCDE that couples the $1e$-$1h$ channel of the 2-GF with the $2e$-$2h$ channel of the 4-GF will be denoted $(4,0)$-MCDE

Moreover we have to approximate the multichannel self-energy.
We choose to approximate each term in the multichannel self-energy by including all contributions that are first order in the interaction. This corresponds to correlating pairs of particles at the RPA+exchange (RPAx) level.
In Ref.~\onlinecite{riva_prl} we have explicitly given the multichannel self-energy for the coupling of the 1-GF and the 3-GF. In the next subsections we will discuss in detail the MCDE that couples the $1e$-$1h$ channel of the 2-GF with the $2e$-$2h$ channel of the 4-GF as an example of coupling even-order Green's functions.
Other couplings of odd or even GFs can be deduced from these two examples.

%
\subsection{The 4-body Green's function}\label{G4:Sec}
%
The 4-GF is defined by 
%
\begin{align}\label{G4def:eq}
    & G_4(1,2,3,4,1',2',3',4') = 
    \nonumber \\ & \langle\Psi_0 ^N|\hat{T}[\hat{\psi}(1)\hat{\psi}(2)\hat{\psi}(3)\hat{\psi}(4)\hat{\psi}^\dag(4')\hat{\psi}^\dag(3')\hat{\psi}^\dag(2')\hat{\psi}^\dag(1')]|\Psi_0 ^N\rangle,
\end{align}
with $1=(x_1,t_1)$, where $x_1=(\mathbf{r}_1,\sigma_1)$, $|\Psi_0^N\rangle$ the ground-state many-body wavefunction of an $N$-electron system, and $\hat{T}$ the time ordering operator defined as

\begin{align}
\hat{T}[\hat \psi(1)...\hat \psi(n)]&=\sum_p (-1)^p  \theta(t_{p_1}>t_{p_2})...\theta(t_{p_{n-1}}>t_{p_n}) \nonumber\\
&\times\hat \psi(p_1)...\hat \psi(p_n),
\end{align}
where the sum runs over all the possible permutations, and $p$ is the number of permutations with respect to the initial order $t_1...t_n$.
The field operators in the Heisenberg picture are defined as
\begin{equation}\label{Heisenberg:eq}
    \hat \psi(1)=\hat U^{\dagger}(t_1,0)\hat \psi(x_1)\hat U(t_1,0)=e^{i\hat H t_1} \hat \psi(x_1) e^{-i\hat H t_1},
\end{equation}
where $\hat{U}(t,t')$ is the time evolution operator and we have assumed that the Hamiltonian $\hat H$ is time-independent.

The 4-GF depends on eight times or seven time differences when the Hamiltonian is time-independent. The total number of permutations of the field operators in Eq.~\eqref{G4def:eq} due to the $\hat{T}$ operator is $8!=40320$.  
Different choices of the time ordering yield different orders of the field operators and, therefore, different physical information. The 4-GF describes the propagation of four particles (electrons or holes) and it can be split into five components: $G_4^{4e}$, $G_4^{3e-1h}$, $G_4^{2e-2h}$, $G_4^{1e-3h}$ and $G_4^{4h}$. 

As mentioned above, in this work we focus on processes with a final state that have the same number of electrons as the initial ground state ($M=N$). For this reason, in the following we analyze only the $2e-2h$ component of the 4-GF.
We will show that by coupling it to the $1e$-$1h$ channel of the 2-GF we can treat single and double neutral excitations on equal footing.
In order to isolate the $2e$-$2h$ channel of the 4-GF we choose the following time differences 
\begin{align}\label{timesG4:eq}
   && \tau_{12}=0^+, && \tau_{22'}=0^+, && \tau_{2'1'}=0^+,&& \tau_{34}=0^+,\nonumber\\  && \tau_{44'}=0^+, && \tau_{4'3'}=0^+, && \tau=t_1-t_3,
\end{align}
where $\tau_{ij}=t_i-t_j$.
The time difference $\tau$ corresponds to the simultaneous propagation of the two electrons and two holes in the system. The other time differences vanish which corresponds to the simultaneous creation (and destruction) of 
the 4 particles.
We thus obtain the following expression for $G_4^{2e-2h}$,
\begin{align}\label{G4ehdef:eq}
    & G_4^{2e-2h}(x_1,x_2,x_3,x_4,x_{1'},x_{2'},x_{3'},x_{4'};\tau) =
    \nonumber \\ & \langle\Psi_0 ^N|T[(\hat{\psi}(x_1)\hat{\psi}(x_2)\hat{\psi}^{\dagger}(x_{2'})\hat{\psi}^{\dagger}(x_{1'}))_{t_1}
    \nonumber \\ & \times
    (\hat{\psi}(x_3)\hat{\psi}(x_4)\hat{\psi}^\dag(x_{4'})\hat{\psi}^\dag(x_{3'}))_{t_3}]|\Psi_0 ^N\rangle,
\end{align}
where the subscripts $t_1$ and $t_3$ on the right-hand side of Eq.~\eqref{G4ehdef:eq} indicate that all the operators in the brackets act at time $t_1$ and $t_3$, respectively.
By introducing the closure relation in Fock space $\sum_M \sum_k | \Psi_k^{M}\rangle \langle\Psi_k^{M} | =\mathbb I$ in Eq.~\eqref{G4ehdef:eq}, 
where $|\Psi_k^{M}\rangle$ indicates the $k$-th eigenstate of the $M$-electron system, and Fourier transforming with respect to $\tau$ leads to the following spectral representation
\begin{align}\label{G4spectral:eq}
        &G_4^{2e-2h}(x_1,x_2,x_3,x_4,x_{1'},x_{2'},x_{3'},x_{4'};\omega) =
        \nonumber \\ & i\lim_{\eta \to 0^+} \sum_n \big[ \frac{X_n(x_{1},x_2,x_{1'},x_{2'})\Tilde{X}_n(x_{3},x_{4},x_{3'},x_{4'})}{\omega - (E_n^N - E_0^N) + i\eta}  \nonumber \\ &
        -\frac{\Tilde{X}_n(x_{1},x_2,x_{1'},x_{2'})X_n(x_{3},x_4,x_{3'},x_{4'})}{\omega + (E_n^N - E_0^N) - i\eta}\big],
\end{align}
with 
\begin{align}\label{Eqn:G4_amplitudes}
          X_n(x_{1},x_2,x_{1'},x_{2'})&= \langle\Psi_0 ^N|\hat{\psi}(x_{1})\hat{\psi}(x_2)\hat{\psi}^\dag(x_{2'})\hat{\psi}^\dag(x_{1'})|\Psi_n ^N\rangle , \nonumber \\  \Tilde{X}_n(x_{1},x_{2},x_{1'},x_{2'}) &=\langle\Psi_n ^N|\hat{\psi}(x_{1})\hat{\psi}(x_2)\hat{\psi}^\dag(x_{2'})\hat{\psi}^\dag(x_{1'})|\Psi_0 ^N\rangle,
\end{align}
and $E_n^N$ the eigenvalues of the $N$-electron system.
We see that Eq.~\eqref{G4spectral:eq} has the same poles as the $1e$-$1h$ channel of the 2-GF, i.e., at the neutral excitation energies of the system. 
Instead, the corresponding amplitudes, given
in Eq.~(\ref{Eqn:G4_amplitudes}), are different. 
For this reason, the $2e$-$2h$ channel of the 4-GF includes the $1e$-$1h$ channel of the 2-GF. 
One can indeed verify that
\begin{align}
    &G_2^{1e-1h}(x_2,x_3,x_{2'},x_{3'},\omega) = \dfrac{1}{(N-1)^2}
    \nonumber \\ &
     \times \int dx_1dx_4G_4^{2e-2h}(x_1,x_2,x_3,x_4,x_{1},x_{2'},x_{3'},x_{4},\omega).
\end{align}
%

In practice it is convenient to project the 4-GF onto a basis.
Using the following transformation 
\begin{align}\label{G4changebasis:eq}
       G_{ijln;mokp}^{2e-2h}(\omega)&= \int dx_1dx_2dx_3dx_4dx_{1'}dx_{2'}dx_{3'}dx_{4'}\nonumber\\
       &\times \phi_i^*(x_1)\phi_j^*(x_2)\phi_l(x_{2'})\phi_n(x_{1'})\nonumber\\
    &\times G_4^{2e-2h}(x_1,x_2,x_3,x_4,x_{1'},x_{2'},x_{3'},x_{4'};\omega)\nonumber\\
     &\times\phi_m(x_{4'})\phi_o(x_{3'})\phi_k^*(x_3)\phi_p^*(x_4),
\end{align}
the spectral representation in Eq.~\eqref{G4spectral:eq} can be rewritten as
\begin{align}\label{G4basis:eq}
    G_{ijln;mokp}^{2e-2h}(\omega)&=i\sum_{n'} \frac{X_{n'}^{ijln} \tilde X_{n'}^{mokp}}{\omega-(E_{n'}^{N}-E_0^N)+i\eta}\nonumber\\
    &-i\sum_{n'} \frac{\tilde X_{n'}^{ijln}  X _{n'}^{mokp}}{\omega+(E_{n'}^N-E_0^{N})-i\eta},
\end{align}
in which
\begin{align}    \label{G4basis_e:eq}
    X_{n'}^{ijln}=\langle \Psi_0^N|\hat c_i \hat c_j \hat c^{\dagger}_l c^{\dagger}_n |\Psi_{n'}^{N}\rangle &&
    \tilde X_{n'}^{mokp}=\langle \Psi_{n'}^{N}|\hat c_k \hat c_p \hat c^{\dagger}_m \hat c^{\dagger}_o    |\Psi_0^N\rangle,
\end{align}
where $\hat{c}^{\dagger}$ and $\hat{c}$ are creation and annihilation operators, respectively. In the following we will drop the superscript $2e-2h$ for notational convenience.
\subsection{Independent-particle 4-GF}
%
Let us analyse the IP 4-GF which we need to solve the MCDE. Using Wick's theorem we can write it as the following determinant
\begin{align}\label{G04wick:eq}  
&G^0_{4}(1,2,3,4,1',2',3',4') = 
\nonumber \\
&\begin{vmatrix}
    G^0_1(1,1') & G^0_1(2,1') & G^0_1(3,1') & G^0_1(4,1')\\
    G^0_1(1,2') & G^0_1(2,2') & G^0_1(3,2') & G^0_1(4,2')\\
    G^0_1(1,3') & G^0_1(2,3') & G^0_1(3,3') & G^0_1(4,3')\\
    G^0_1(1,4') & G^0_1(2,4') & G^0_1(3,4') & G^0_1(4,4')\\
    \end{vmatrix},
\end{align}
which generates $4!=24$ terms, each composed of a product of four IP 1-GF.
Using the time diferences given in Eq.~\eqref{timesG4:eq} we can distinguish three  cases with each case having a different number of IP 1-GFs in the product that depend on the time difference $\tau = t_1 - t_3$. 
We can have either four, two, or zero of those IP 1-GFs. They correspond to the propagation of four, two, or zero independent particles, respectively.
The latter contribution can be neglected since the corresponding spectral function has no poles.
More details are given in appendix \ref{App:NI_G4}. 
Here we report an example of the propagation of two particles and of the propagation of four particles.  The spectral representations of these two contributions to $G^0_{ijln;mokp}$ are given by

\begin{align}
&-G^0_{i;n}G^0_{p;m}[G^0_{j;o}G^0_{k;l}](\omega)=i\frac{\delta_{in}\delta_{pm}\delta_{jo}\delta_{kl}(1-f_i)(1-f_p)f_{lj}}{\omega-
\Delta\epsilon_{jl}+i\eta \text{sign}f_{lj}}\label{G02multi:eq}\\
&[G^0_{i;m}G^0_{j;o}G^0_{k;l}G^0_{p;n}](\omega)=- i
\frac{\delta_{im}\delta_{jo}\delta_{lk}\delta_{np}f_{ni}f_{li}f_{nj}}{\omega- \Delta\epsilon_{in} - \Delta\epsilon_{jl}+i\eta \text{sign}f_{lj}},\label{G04multi:eq}
\end{align}
respectively, where  $[G^0_{i;j}..G^0_{k;l}](\omega)$ implies a frequency convolution, and $G^0_{i;l}=-i(1-f_i)\delta_{il}$.
%
%
%
%
We note that the occupation numbers $f_{in}f_{il}f_{jn}$ in Eq.~\eqref{G04multi:eq} restrict this contribution to $G_{4}^0$ to its $2e2h$ and $2h2e$ channels. The indices $i,j,m,o$ refer to conduction (valence) states and they describe the $2e$ ($2h$) propagation, while $l,n,k,p$ refer to valence (conduction) states and they describe the $2h$ ($2e$) propagation. 
While the poles of Eq.~\eqref{G02multi:eq} contain energy differences of a conduction and valence state, the poles of Eq.~\eqref{G04multi:eq} contain a sum of two eigenenergy differences of a conduction and valence state, which represent approximate double electron-hole excitations already at the noninteracting level.
We will therefore refer to these terms as single and double excitation contributions, respectively.
These contributions do not couple in $G^0_4$, i.e., in its matrix representation $G^0_{ijln;mokp}$, $G^0_{4}$ is block diagonal with a single excitation and a double excitation block.

It can be verified that $G^0_{ijln;mokp}$  is invariant upon the permutation of the indices in any or several of the couples ($i,j$), ($l,n$), ($m,o$), and ($k,p$) (modulo a minus sign for an odd number of permutations).\footnote{We note that the symmetry in the permutation of the indices is also fulfilled by the interacting $G_4$ as can be seen from Eq.~\eqref{G4basis:eq}.}
Moreover, also the following symmetry relation holds

\begin{align}
    \label{G4symmetry:eq}
    -G^{0,4\text p}_{cjlc;c'okc'}(\omega) &= G^{0,2\text p}_{jl;ok}(\omega) \quad (j\neq c, k \neq c' \quad \forall\,c,c')
\end{align}
where $c$ and $c'$ refer to conduction states, i.e., $f_c = f_{c'} = 0$, and $G^{0,2\text p}_{jl;ok}(\omega) $ is defined by Eq.~\eqref{G0n2p:eq}.
As a consequence of the above symmetries $G^0_4(\omega)$ is singular as it contains redundant contributions.
It can be verified that upon elimination of these superfluous contributions one obtains Eq.~\eqref{Eqn:G0n}.

It is convenient to define the four-body correlation function ${L}^0_{4}=iG^0_{4}$. 
With all the restrictions over the space discussed above, $L^0_{4}$ reads  
\begin{align}\label{L04matrix:eq}
  {L}^0_{4}(\omega) =
    \begin{pmatrix}
    {L}^{\text{0,2p}}(\omega) & 0\\
    0 &  {L}^{\text{0,4p}}(\omega)
    \end{pmatrix},
\end{align}
with
\begin{align}
&{L}^{\text{0,2p}}_{jl;ok}(\omega)= \frac{\delta_{jo}\delta_{lk}f_{jl}}{\Delta\epsilon_{jl} -\omega +i\eta \text{sign}f_{jl}},
\\
& {L}^{\text{0,4p}}_{i>jl>n;m>ok>p}(\omega)= 
\frac{\delta_{im}\delta_{jo}\delta_{lk}\delta_{np}f_{in}f_{il}f_{jn}}{ \Delta\epsilon_{in} + \Delta\epsilon_{jl}-\omega+i\eta \text{sign}f_{jl}} .\label{L04multi:eq}
\end{align} 
Now that we defined the IP 4-GF in the $2e$-$2h$ channel it remains to find an approximation to the 4-body self-energy. This will then allow us to solve the $(4,0)$-MCDE for $G^{2e-2h}_4(\omega)$.

All the relations we have obtained hold for any 4-GF built from IP 1-GFs. 
In particular, it is convenient to use the Hartree-Fock 1-GF as the IP 1-GF.
Therefore, in the following, it will be understood that $G^0_{1}$ refers to the Hartree-Fock 1-GF.

\subsection{$(4,0)$-MCDE}\label{multiG2G4:sec}
The $(4,0)$-MCDE is obtained from Eq.~\eqref{multiDyson:eq} by setting $n=4$ and by considering a static self-energy.
By defining ${L}_{4}=iG^{2e-2h}_{4}$, we obtain
\begin{equation}\label{multi_Dyson4:eq}
    L_4(\omega)=L^0_{4}(\omega)+ L^0_{4}(\omega) \tilde{\Sigma}_4  L_4(\omega)
\end{equation}
where $\tilde{\Sigma}_4=-i\Sigma_4$ is the multichannel self-energy. 
It is defined by
\begin{equation}\label{self4:eq}
    \tilde{\Sigma}_4=\begin{pmatrix}
        \tilde{\Sigma}^{2\text p} & \tilde{\Sigma}^\text{2p/4p}\\
        \tilde{\Sigma}^\text{4p/2p} & \tilde{\Sigma}^{4\text p}
    \end{pmatrix}.
\end{equation}
The approximation that we use for it is analogous to the approximation we have used for the $(3,1)$-MCDE.~\cite{riva_prl,riva_prb}
We include all contributions in $\tilde{\Sigma}_4$ that are first order in the interaction.
This corresponds to letting each pair of particles interact at the RPAx level, i.e., through a direct and an exchange interaction. 
The head of the matrix in Eq.~\eqref{self4:eq} thus corresponds to the standard RPAx kernel of the BSE, i.e. 
\begin{equation}\label{XiRPAtilde:eq}
\tilde{\Sigma}^{2\text p}_{jl;ok}=\bar{v}_{jkol}
\end{equation}
where $\bar v_{jkol}=v_{jkol}-v_{jklo}$ with
\begin{equation}
    v_{jkol}=\int dx_1 dx_2 \phi^*_j(x_1)\phi^*_k(x_2)v(\mathbf{r}_1,\mathbf{r}_2)\phi_o(x_2)\phi_l(x_1),\label{potential:eq}
\end{equation}
%
For the other three  components of $\tilde{\Sigma}_4$ our approximation yields the following static expressions
\begin{align}
  \tilde{\Sigma}^{\text{4p}}_{ijln;mokp}&=\delta_{im}\delta_{jo}\bar{v}_{pknl}+\delta_{lk}\delta_{np}\bar{v}_{ijmo}\nonumber\\
    &+\delta_{lp}\delta_{nk}\bar{v}_{ijom}+\delta_{io}\delta_{jm}\bar{v}_{pkln}
    \nonumber\\
    &+\delta_{im}\delta_{np}\bar{v}_{jklo}+\delta_{im}\delta_{lk}\bar{v}_{jpno}\nonumber\\
    &+\delta_{jo}\delta_{np}\bar{v}_{iklm}+\delta_{jo}\delta_{lk}\bar{v}_{ipnm}\nonumber\\
    &-\delta_{im}\delta_{nk}\bar{v}_{jplo}-\delta_{im}\delta_{lp}\bar{v}_{jkno}\nonumber\\
    &- \delta_{jo}\delta_{nk}\bar{v}_{iplm}-\delta_{jo}\delta_{lp}\bar{v}_{iknm} \nonumber \\
    &-\delta_{io}\delta_{np}\bar{v}_{jklm}- \delta_{jm}\delta_{np}\bar{v}_{iklo}\nonumber\\
    &-\delta_{io}\delta_{lk}\bar{v}_{jpnm}-\delta_{jm}\delta_{lk}\bar{v}_{ipno}
    \nonumber\\
    &+\delta_{io}\delta_{lp}\bar{v}_{jknm}+\delta_{io}\delta_{nk}\bar{v}_{jplm}\nonumber\\
    & +\delta_{jm}\delta_{lp}\bar{v}_{ikno}+\delta_{jm}\delta_{nk}\bar{v}_{iplo} \label{selfbody4:eq}
  \end{align}
\begin{align}
    \tilde{ \Sigma}^\text{2p/4p}_{jl;mokp} &=\delta_{jo}\bar{v}_{kpml}+  \delta_{lk}\bar{v}_{jpom}+ \delta_{lp}\bar{v}_{jkmo}  + \delta_{jm}\bar{v}_{pkol}\label{selfcoupling4:eq}\\
    \tilde{ \Sigma}^\text{4p/2p}_{ijln;ok}&= \delta_{jo}\bar{v}_{ikln} + \delta_{lk}\bar{v}_{ijno}+\delta_{nk}\bar{v}_{ijol}+\delta_{io}\bar{v}_{jknl} \label{selftildecoupling4:eq}
\end{align}
%
The first four terms of $\tilde{\Sigma}^{4\text p}$ given on the right-hand side of Eq.~\eqref{selfbody4:eq} account for all the electron-electron and hole-hole interactions, whereas the other terms describe all the one-electron-one-hole interactions. Finally, we note that the multichannel self-energy is hermitian. 
\subsection{Diagrammatic analysis}
It is instructive to represent the $(4,0)$-MCDE~\eqref{multi_Dyson4:eq} diagrammatically according to 
\begin{widetext}
\begin{equation}\label{Eqn:Dyson_diag}
    \begin{gathered}
        \includegraphics[width=1\textwidth,clip=]{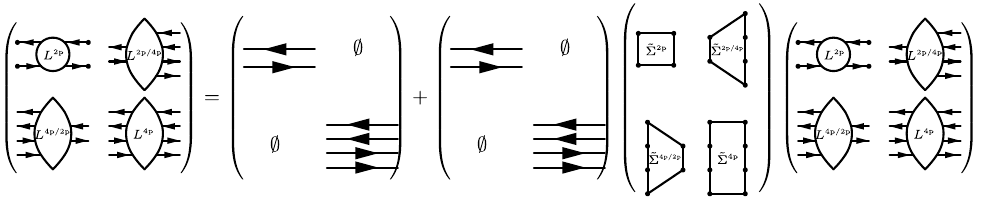}
    \end{gathered}
\end{equation}.
\end{widetext}
From Eq.~(\ref{Eqn:Dyson_diag}), it is evident that $L_4$ consists of the standard electron-hole $L^{2\text{p}}$  (\raisebox{-0.3\totalheight}{\includegraphics[width=0.03\textwidth]{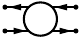}}) along with an explicit four-body component $L^{4\text p}$ (\raisebox{-0.3\totalheight}{\includegraphics[width=0.03\textwidth]{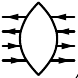}}), and the coupling between the $L^{2\text{p}}$ and $L^{4\text p}$, denoted as $L^{2\text p/4\text p}$ (\raisebox{-0.3\totalheight}{\includegraphics[width=0.03\textwidth]{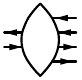}}) and $L^{4\text p/2\text p}$ (\raisebox{-0.3\totalheight}{\includegraphics[width=0.03\textwidth]{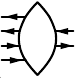}}). We represent the self-energy coupling terms, $\tilde{\Sigma}^{2\text p/4\text p}_{jl;mokp}$ and $\tilde{\Sigma}^{4\text p/2\text p}_{ijln;ok }$, and the body of the self-energy, $\tilde{\Sigma}^{4\text p}_{ijln;mokp}$,  using isosceles trapezoids and a rectangle, respectively, to reflect their dimensions.

\begin{figure}
\centerline{
\includegraphics[width=0.4\textwidth,clip=]{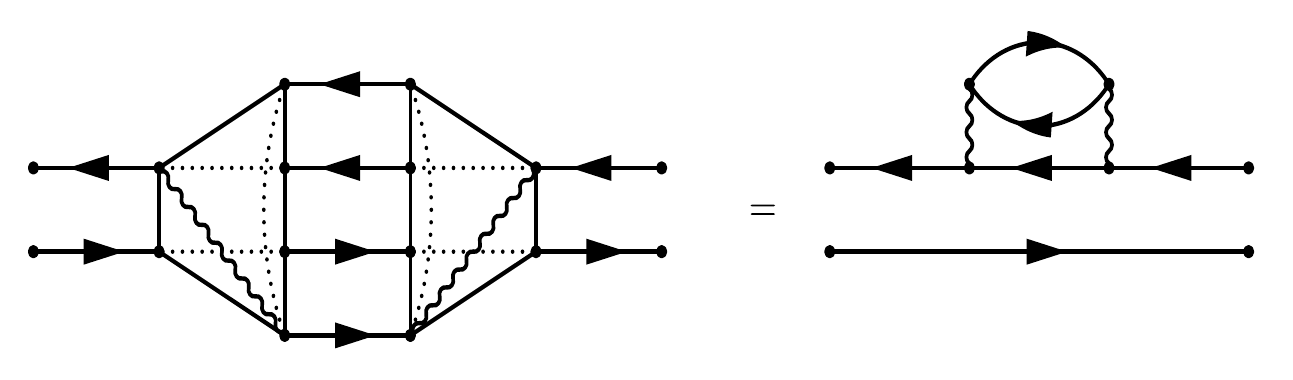}\hfill
}
\caption{\label{second4_oneline:fig} An example of a combination of multichannel self-energy coupling terms that dress a single $G^0_1$ line.}
\end{figure}

\begin{figure}
\centerline{
\includegraphics[width=0.4\textwidth,clip=]{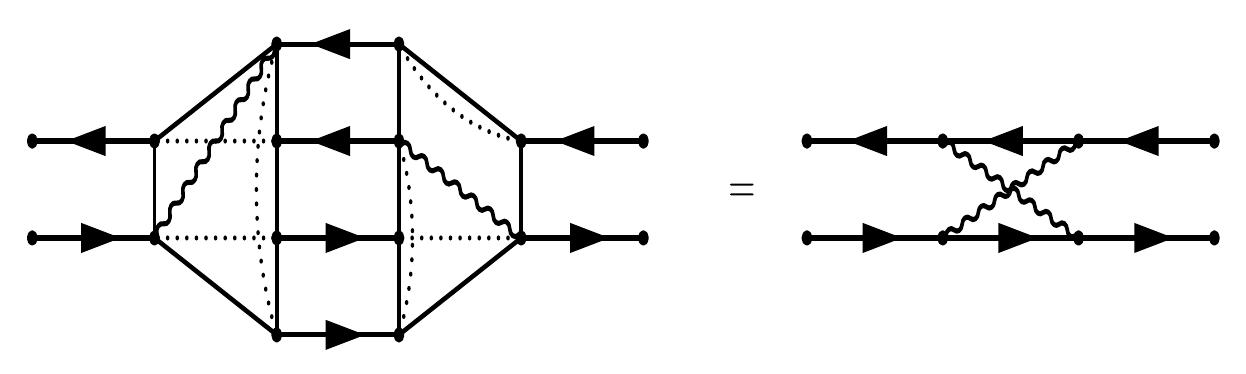}\hfill
}
\caption{\label{second4_cross:fig} An example of a combination of multichannel self-energy coupling terms that dress the interaction between two particles. Here they create a diagram where the two interaction lines cross each other.}
\end{figure}

\begin{figure}
\centerline{
\includegraphics[width=0.4\textwidth,clip=]{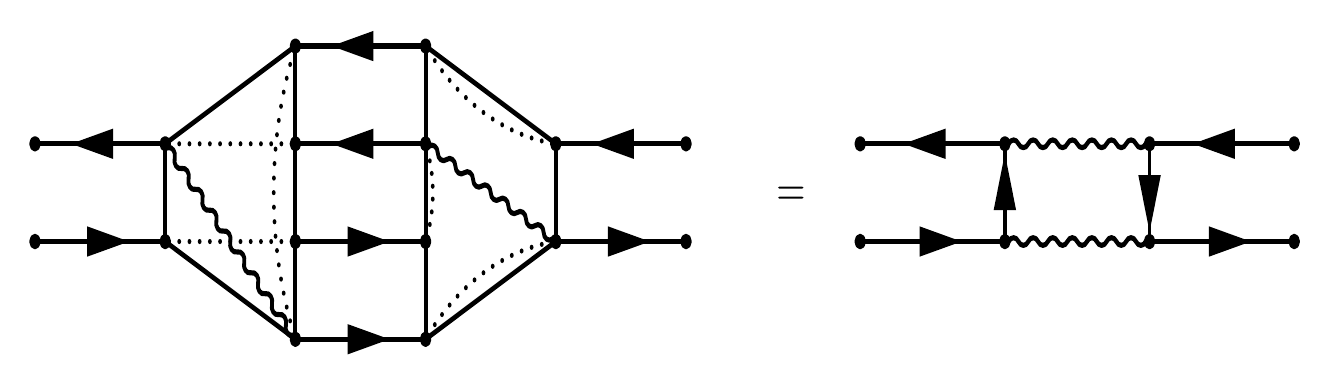}\hfill
}
\caption{\label{second4_ladder:fig} An example of a combination of multichannel self-energy coupling terms that dress the interaction between the two particles. Here they create a ladder diagram.}
\end{figure}

\begin{figure}
\centerline{
\includegraphics[width=0.4\textwidth,clip=]{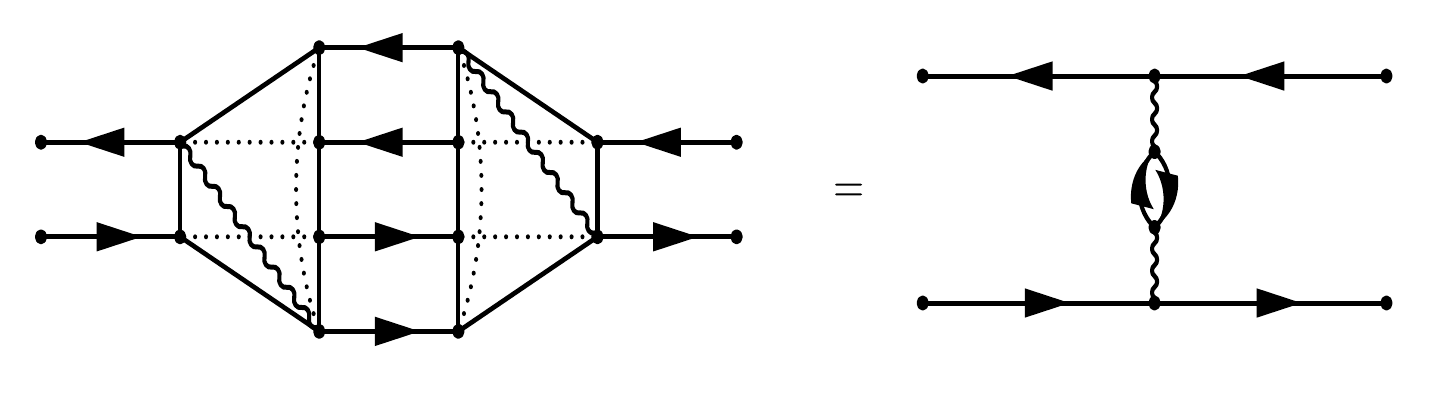}\hfill
}
\caption{\label{second4_screening:fig} An example of a combination of multichannel self-energy coupling terms that dresses the interaction between the two particles. Here they screened an $eh$ attraction.}
\end{figure}

\begin{figure}
\centerline{
\includegraphics[width=0.4\textwidth,clip=]{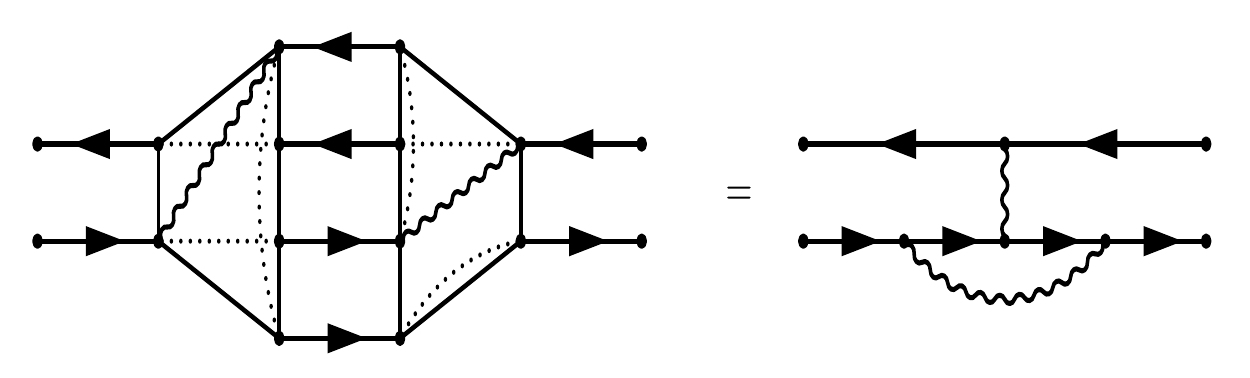}\hfill
}
\caption{\label{second4_attraction_vertex:fig} An example of a combination of multichannel self-energy coupling terms that creates a mixed term. Here, the $eh$ interaction is between a hole and an electron that is already dressed.}
\end{figure}

To represent diagrammatically the multichannel self-energy in Eqs.~\eqref{selfbody4:eq}-\eqref{selftildecoupling4:eq}, it is convenient to first rewrite them in real space. Those real-space expressions can be obtained from expressions similar to Eq.~\eqref{G4changebasis:eq} and they are given in appendix \ref{sec:realspace}.
The expressions of the $(4,0)$-MCDE in real space makes it easier to understand the diagrammatic structure. 
The head $\tilde{\Sigma}^{2p}$ is the kernel of the $1e$-$1h$ BSE in the RPAx approximation. 
Diagrammatically it can be represented as
\begin{equation}\label{ppBSE-RPAx}
    \begin{gathered}
        \includegraphics[width=0.47\textwidth,clip=]{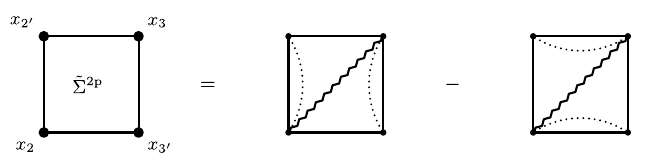},
    \end{gathered}
\end{equation}
where a dotted line represent a Dirac delta function and a wiggly line represents the bare Coulomb interaction.
The body $\tilde{\Sigma}^{4\text p}$~\eqref{selfbody4:eq} is given by 
\begin{widetext}
\begin{align}
      \begin{gathered}
        \includegraphics[width=1\textwidth,clip=]{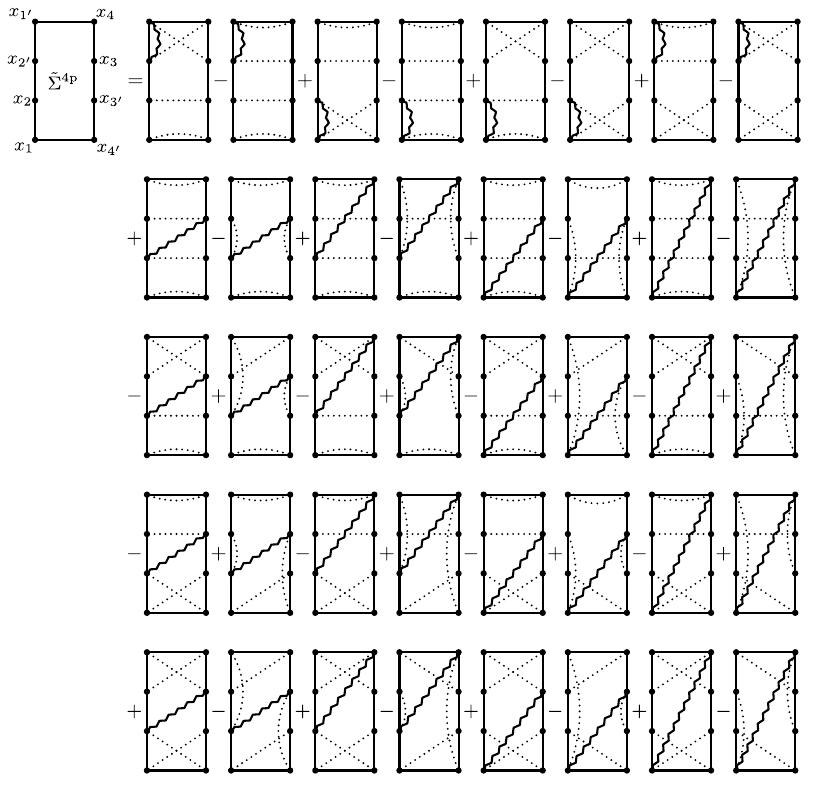}\label{self4_body:fig}
    \end{gathered}.
\end{align}
\end{widetext}
The contributions in the first row on the right-hand side of Eq.~(\ref{self4_body:fig}) account for the particle-particle interaction, while the remaining contributions are needed to account for
all electron-hole interactions.
Finally, the coupling terms $\tilde{\Sigma}^{2p/4p}$ and $\tilde{\Sigma}^{4p/2p}$ in Eqs.~\eqref{self4_couplingreal:eq} and~\eqref{self4_couplingtildereal:eq}, respectively, are represented by
\begin{widetext}
\begin{align}
    \begin{gathered}
        \includegraphics[width=0.95\textwidth,clip=]{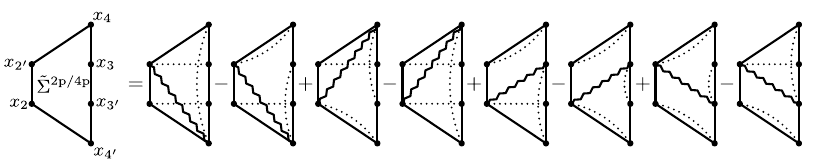}
    \end{gathered} \label{self4_coupling:fig},\\
    \begin{gathered}
        \includegraphics[width=0.95\textwidth,clip=]{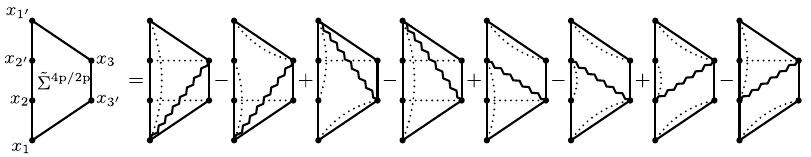}
    \end{gathered}  \label{self4_coupling_tilde:fig}.
\end{align}
\end{widetext}
To understand which diagrams are added to those included in $L^{0,\text{2p}}$ when solving the $(4,0)$-MCDE, one can iterate Eq.~\eqref{Eqn:Dyson_diag} and inspect the head of the final matrix which corresponds to $L^{\text{2p}}(\omega)$.
The first iteration is equivalent to the first iteration of the BSE with the RPAx kernel. 
Upon a second iteration, the coupling terms $\Sigma^{2p/4p}$ and  $\Sigma^{4p/2p}$ will contribute to the head of $L_4$. 
They add correlation in three different ways: 1) they dress single $G^0_{1}$ lines (see Fig.~\ref{second4_oneline:fig} for an example). Due to this, the independent-particle two-body term $L^{0,\text{2p}}$ in Eq.~\eqref{L04multi:eq}  
should not be evaluated beyond the Hartree-Fock level as this would lead to a double counting of diagrams and, therefore, of correlation;
2) they dress the interaction between the two particles, for example by screening the interaction (see Figs.~\ref{second4_cross:fig}-\ref{second4_screening:fig} for examples); 3) they create mixed terms in which a dressed $G^0_{1}$ interacts with a bare $G^0_{1}$ (see Fig.~\ref{second4_attraction_vertex:fig}).
We note that after two iterations of the $(4,0)$-MCDE with the coupling terms in Eqs.~\eqref{self4_coupling:fig} and~\eqref{self4_coupling_tilde:fig}, we obtain all diagrams that are second-order in the interaction.
In other words, the $(4,0)$-MCDE is exact up to second order in the interaction.

%

Figure~\ref{second4_screening:fig} shows an $eh$ interaction that is screened to first-order in the polarizability $-iG^{0}_1G^0_1$.
Therefore, as for the $(3,1)$-MCDE~\cite{riva_prl,riva_prb}, the screening of the interaction is naturally included in the $(4,0)$-MCDE even though the self-energy only involves the bare Coulomb potential.

\begin{figure}
\centerline{
\includegraphics[width=0.4\textwidth,clip=]{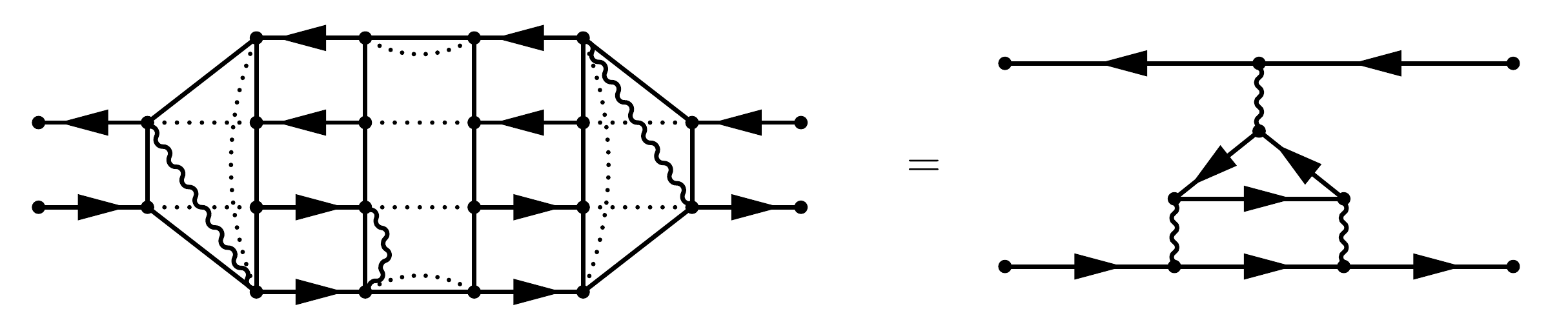}\hfill
}
\caption{\label{third4:fig} An example of a diagram that is of third-order in the interaction.}
\end{figure}

Upon a third iteration of the $(4,0)$-MCDE, also the body of the multichannel self-energy in Eq.~\eqref{self4_body:fig} will contribute to the head of $L_4(\omega)$. This yields a large number of diagrams that are all of third order in the interaction. We report an example in Fig.~\ref{third4:fig}.
Further iterations will yield diagrams of all higher orders in the interaction.
All of these diagrams are naturally included in $L^{\text{2p}}(\omega)$ when solving the $(4,0)$-MCDE.

We note that to further dress the diagrams, one can replace the Hartree-Fock $G^0_{1}$ lines in the body of $L^0_4(\omega)$ by $G^0_{1}$ lines obtained from second Born, $GW$ or T-matrix quasiparticle calculations, for example.
This could be seen as a numerically efficient way to (partially) compensate for the truncation of the full MCDE, i.e., the absence of the coupling to the $3e$-$3h$ channel of the 6-GF, etc.
In practice, it would correspond to replacing the HF eigenenergies in Eq.~\eqref{L04multi:eq} by QP energies obtained from second Born, $GW$ or T-matrix calculations.

We note that diagrams that add correlation to a single $G^0_1$ line are needed to create also other terms. For example, the diagrams in Figs.~\ref{second4_oneline:fig} and~\ref{second4_ladder:fig} are created with the same coupling term. Moreover, as for the (3,1)-MCDE,\cite{riva_prb} one can go beyond the RPAx approximation to $\tilde{\Sigma}_4$ by dressing all the particle-particle interactions (first row) and all the direct electron-hole interactions (first, third, fifth and seventh diagrams of the remaining rows) of Eq.~(\ref{self4_body:fig}), for example, by using a screened Coulomb interaction~\cite{Hed65,Deilmann_2016,Tar18,Torche_2019,Sab22,Sab23}.

\subsection{Effective four-particle Hamiltonian}
In the previous sections we have given details about the derivation of the $(4,0)$-MCDE as well as an analysis of the diagrams that are included when using an RPAx approximation for the corresponding self-energy.
In this section we will show how the $(4,0)$-MCDE can be solved using standard numerical techniques.
Since the MCDE self-energy is static and Hermitian it is convenient to rewrite the $(4,0)$-MCDE in terms of an eigenvalue problem with an effective Hamiltonian. 
Upon solving the eigenvalue problem we can construct $L_4(\omega)$ with the eigenvalues and eigenvectors of the effective Hamiltonian.

Following a similar strategy as in Refs.~\onlinecite{Oni02,riva_prl}, we invert Eq.~(\ref{multi_Dyson4:eq}) 
according to
\begin{equation}
L_4(\omega)=[L^{0,-1}_{4}(\omega)-\tilde{\Sigma}_4]^{-1}.
\end{equation}
Therefore, $L_4(\omega)$ can be written as
\begin{widetext}
\begin{align}\label{L-1}
 L_4(\omega)&=\begin{pmatrix}
       \frac{\delta_{jo}\delta_{lk}\left(\Delta\epsilon_{jl} -\omega\right)}{f_{jl}}- \tilde{\Sigma}^{2p}_{jl;ok} & -\tilde{\Sigma}^\text{2p/4p}_{jl;m>ok>p}\\
       - \tilde{\Sigma}^\text{4p/2p}_{i>jl>n;ok} &  \frac{\delta_{im}\delta_{jo}\delta_{lk}\delta_{np} \left(\Delta\epsilon_{in} + \Delta\epsilon_{jl}-\omega\right)}{f_{in}f_{il}f_{jn}}{}-\tilde{\Sigma}^{4\text p}_ {i>jl>n;m>ok>p}   
    \end{pmatrix}^{-1}.
 \end{align}
Using $[AB]^{-1} = B^{-1}A^{-1}$ we obtain

\begin{equation}\label{Eqn:L4_0}
 L_4(\omega)=\left[H^{\text{eff}}_4(\omega)-\omega\mathbf{1}\right]^{-1}\begin{pmatrix}
   \delta_{jo}\delta_{lp} f_{jl}&0\\
    0&  \delta_{im}\delta_{jo}\delta_{lk}f_{in}f_{il}f_{jn} \end{pmatrix},
 \end{equation}
where the effective Hamiltonian is defined as
\begin{align}
H^{\text{eff}}_4(\omega)=\begin{pmatrix}
     \delta_{jo}\delta_{lk}\Delta\epsilon_{jl}-f_{jl}\tilde{\Sigma}^{2p}_{jl;ok}& -\tilde{\Sigma}^\text{2p/4p}_{jn;m>ok>p}f_{mp}f_{mk}f_{op}\\
       - \tilde{\Sigma}^\text{4p/2p}_{i>jl>n;ok} f_{ok}& \delta_{im}\delta_{jo}\delta_{lk}\delta_{np} \left(\Delta\epsilon_{in} + \Delta\epsilon_{jl}\right)-f_{in}f_{il}f_{jn}\tilde{\Sigma}^{4\text p}_ {i>jl>n;m>ok>p}              
       \end{pmatrix}.
\end{align}
\end{widetext}

The presence of $f_{jl}$ and $f_{in}f_{il}f_{jn}$ in Eq~ (\ref{Eqn:L4_0}) means that only the 
$(jn=\{cv,vc\}; ok=\{c'v',v'c'\})$, 
$(i>j;l>n=\{c>\tilde{c};v>\tilde{v},v>\tilde{v};c>\tilde{c}\}; ,m>o;k>p=\{c'>\tilde{c}';v'>\tilde{v}',v'>\tilde{v}';c'>\tilde{c}'\})$ components of $\left[H^{\text{eff}}_4(\omega)-\omega\mathbf{1}\right]^{-1}$ will contribute to $L_4$. 
Defining the effective Hamiltonian in this space as $\bar{H}_4^{\text{eff}}$, we can now write
\begin{equation}\label{Eqn:eigen_pb}
\bar{H}_4^{\text{eff}} A_{\lambda} = E_{\lambda}A_{\lambda},
\end{equation}
where $E_{\lambda}$ and $A_{\lambda}$ are the eigenvectors and eigenvalues of $\bar{H}_4^{\text{eff}}$, respectively.
The spectral representation of $\left[\bar{H}_4^{\text{eff}}-\mathbf{1}\omega  \right]^{-1}$ can thus be written as
%
\begin{equation}
\left[ \bar{H}_4^{\text{eff}} -\mathbf{1}\omega\right]^{-1}_{\mu\nu} = \sum_{\lambda} \frac{A^\mu_{\lambda}A^{*\nu}_{\lambda}}{E_{\lambda}-\omega}.
\end{equation}
We thus obtain the following expression for $L_4$,
%
\begin{equation}
L_4 (\omega) =
\begin{pmatrix}
        L^{2\text{p}}(\omega) & L^{\text{2p/4p}}(\omega)  \\
        L^{\text{4p/2p}}(\omega)  & L^{4\text p}(\omega)
    \end{pmatrix},
\end{equation}
with
\begin{align}
    L^{\text{2p}}_{jl;ok}(\omega)& =  \sum_{\lambda}\frac{A^{jl}_{\lambda}A^{*ok}_{\lambda}}{E_{\lambda}-\omega}\label{Eqn:L},
    \\
    L^{\text{2p/4p}}_{jl;m>ok>p}(\omega)&=  \sum_{\lambda}\frac{A^{jl}_{\lambda}A^{*mokp}_{\lambda}}{E_{\lambda}-\omega},
    \\
    L^{\text{4p/2p}}_{i>jl>n;ok}(\omega)& =  \sum_{\lambda}\frac{A^{ijln}_{\lambda}A^{*ok}_{\lambda }}{E_{\lambda}-\omega},
    \\
    L^{\text{4p}}_{i>jl>n;m>ok>p}(\omega)&=  \sum_{\lambda}\frac{A^{ijln}_{\lambda}A^{*mokp}_{\lambda}}{E_{\lambda}-\omega}.
\end{align}
Equation (\ref{Eqn:L}) is the equation of interest here, since it is linked to neutral excitations and absorption spectra.
The eigenvalue problem in Eq.~\eqref{Eqn:eigen_pb} can be solved by direct diagonalisation or by more efficient iterative techniques.
Using the Haydock-Lanczos solver~\cite{Hay72,Schm03,Her05} we can directly solve for the spectral function.
The numerical scaling of the calculation will then be determined by the construction of $\bar{H}_4^{\text{eff}} $ 
which scales as $N_v^3 N^3_c$, where $N_v$ and $N_c$ are the number of occupied and unoccupied states, respectively.
The overall scaling with respect to the number of electrons $N$ is thus $N^6$.
%
%
\section{Illustration on the two-level Helium-like model\label{sec:applications}}
As an illustration of the $(4,0)$-MCDE we focus on double excitations, which are currently difficult to capture within the standard approximations to the BSE. \cite{Stri88,Rom09-2,San11,Authier_2020,Loos_2020,Bintrim_2022,Monino_2023}
Although higher-order excitations are usually dark in the experiments, they can play an important role for the description of ground and optically active excited states.\cite{Serrano_1993,Cave_1988,Lappe_2000,Wanko_2005,Boggio_2004,STARCKE200639,CAVE200439,Catalan_2006,HUIXROTLLANT2011120,Kossoski_2024} For example, the lowest-lying singlet state of closed-shell molecules such as polyenes is known to have a HOMO$^2\rightarrow$LUMO$^2$ character. 

Here we employ the two-level two-electron model used in Ref.~\onlinecite{Rom09-2}, where this model is used to study double excitations within the standard BSE with a frequency-dependent $GW$-based kernel. We can compare the eigenvalues of the $(4,0)$-MCDE for this Helium-like model with the eigenvalues of the exact Hamiltonian given by
\begin{equation}
\hat{H}=\sum^2_{i=1}\hat{h}(x_i)+\frac{1}{2}\sum^2_{i, j=1, i\neq j} v(x_i,x_j),
\end{equation}
where the terms on the right-hand side are the noninteracting
Hamiltonian and the two-electron Coulomb interaction, respectively. 
In this model the occupied bonding orbital $\phi_v$ and the virtual antibonding orbital $\phi_c$ are given by the hydrogenic $1s$ and $2s$ wave functions, respectively. Therefore the two-electron repulsion integrals $v_{ijkl}$ can be evaluated analytically, whereas the one-electron energies $\epsilon_i = h_{ii}$ are computed from the Bohr model, i.e., $\epsilon_n= - 13.6 Z^2/n^2$ eV, with $Z$ the
atomic number and $n$ the principal quantum number. 
More details of the Helium-like model can be found in Ref.~\onlinecite{Rom09-2}.

To build the $(4,0)$-MCDE we first have to calculate the HF 1-GF which, in the bonding/antibonding ($\phi_v/\phi_c$) basis, is given by
%
\begin{equation}
	[G^{\text{0}}_1]^{-1}=\left(
	\begin{array}{cc}
		\omega-\epsilon^0_v-\Sigma^{\text{HF}}_{v;v} & -\Sigma^{\text{HF}}_{v;c}  \\
		-\Sigma^{\text{HF}}_{c;v} & \omega-\epsilon^0_c-\Sigma^{\text{HF}}_{c;c}
	\end{array}
	\right).
\end{equation}
%
The effective Hamiltonian $\bar{H}^{\text{eff}}_4$ is built in the basis in which $L^{\text{0}}_4$ is diagonal, which is also the basis in which $G_1^{\text{0}}$ is diagonal. This basis reads $\phi_{\tilde{v}}=\alpha \phi_v+\beta \phi_c$ and $\phi_{\tilde{c}}=\beta \phi_v-\alpha \phi_c$, where the greek letters are the coefficients of the linear combination of the old $\phi_v/\phi_c$ basis, after normalization, which, after convergence, take the values $\{\alpha,\beta\}=\{-0.98, 0.19\}$. The converged HF bonding and antibonding energies read $\epsilon_{\tilde{v}}^{0}=-23.93$ eV and $\epsilon_{\tilde{c}}^{0}=8.59$ eV, respectively. The Hamiltonian $\bar{H}^{\text{eff}}_4$ is hence represented by a 10$\times$10 matrix in the basis $\{\tilde{v}\sigma,\tilde{c}\sigma'\}$, $\{\tilde{c}\sigma,\tilde{v}\sigma'\}$,$\{\tilde{v}\uparrow,\tilde{v}\downarrow,\tilde{c}\uparrow,\tilde{c}\downarrow\}$, $\{\tilde{c}\uparrow,\tilde{c}\downarrow,\tilde{v}\uparrow,\tilde{v}\downarrow\}$.
In this basis, the MCDE self-energy $\Sigma_4$ has the following non-zero elements

\begin{align}
&\tilde{\Sigma}^{2p}_{\tilde{v}\sigma \tilde{c}\sigma';\tilde{v}\tilde{\sigma} \tilde{c}\tilde{\sigma}'}=v_{\tilde{v}\tilde{c}\tilde{v}\tilde{c}}\delta_{\sigma\sigma'}\delta_{\tilde{\sigma}\tilde{\sigma}'}-v_{\tilde{c}\tilde{v}\tilde{v}\tilde{c}}\delta_{\sigma\tilde{\sigma}} \delta_{\sigma'\tilde{\sigma}'},  \\
&\tilde{\Sigma}^{2p}_{\tilde{v}\sigma \tilde{c}\sigma';\tilde{c}\tilde{\sigma} \tilde{v}\tilde{\sigma}'}=v_{\tilde{v}\tilde{v}\tilde{c}\tilde{c}}\delta_{\sigma\sigma'}\delta_{\tilde{\sigma}\tilde{\sigma}'}-v_{\tilde{v}\tilde{v}\tilde{c}\tilde{c}}\delta_{\sigma\tilde{\sigma}} \delta_{\sigma'\tilde{\sigma}'} ,  \\
&\tilde{\Sigma}^{2p}_{\tilde{c}\sigma \tilde{v}\sigma';\tilde{v}\tilde{\sigma} \tilde{c}\tilde{\sigma}'}=-[\tilde{\Sigma}^{2p}_{\tilde{v}\sigma \tilde{c}\sigma';\tilde{c}\tilde{\sigma} \tilde{v}\tilde{\sigma}'}]^* , \\
&\tilde{\Sigma}^{2p}_{\tilde{c}\sigma \tilde{v}\sigma';\tilde{c}\tilde{\sigma} \tilde{v}\tilde{\sigma}'}=-[\tilde{\Sigma}^{2p}_{\tilde{v}\sigma \tilde{c}\sigma';\tilde{v}\tilde{\sigma} \tilde{c}\tilde{\sigma}'}]^* , \\
&\tilde{\Sigma}^{2p/4p}_{\tilde{v}\sigma \tilde{c}\sigma';\tilde{v}\uparrow\tilde{v}\downarrow\tilde{c}\uparrow\tilde{c}\downarrow}=v_{\tilde{v}\tilde{c}\tilde{v}\tilde{v}}\delta_{\sigma\sigma'}-v_{\tilde{c}\tilde{v}\tilde{c}\tilde{c}}\delta_{\sigma\sigma'},  \\
&\tilde{\Sigma}^{2p/4p}_{\tilde{c}\sigma \tilde{v}\sigma';\tilde{c}\uparrow\tilde{c}\downarrow\tilde{v}\uparrow\tilde{v}\downarrow}=\tilde{\Sigma}^{2p/4p}_{\tilde{v}\sigma \tilde{c}\sigma';\tilde{v}\uparrow\tilde{v}\downarrow\tilde{c}\uparrow\tilde{c}\downarrow},
\\
&\tilde{\Sigma}^{4p/2p}_{\tilde{v}\uparrow\tilde{v}\downarrow\tilde{c}\uparrow\tilde{c}\downarrow;\tilde{v}\sigma \tilde{c}\sigma'}=\tilde{\Sigma}^{2p/4p}_{\tilde{v}\sigma \tilde{c}\sigma';\tilde{v}\uparrow\tilde{v}\downarrow\tilde{c}\uparrow\tilde{c}\downarrow}, \\
&\tilde{\Sigma}^{4p/2p}_{\tilde{c}\uparrow\tilde{c}\downarrow\tilde{v}\uparrow\tilde{v}\downarrow;\tilde{c}\sigma \tilde{v}\sigma'}=\tilde{\Sigma}^{2p/4p}_{\tilde{c}\sigma \tilde{v}\sigma';\tilde{c}\uparrow\tilde{c}\downarrow\tilde{v}\uparrow\tilde{v}\downarrow},
 \\
 &\tilde{\Sigma}^{4p}_{\tilde{v}\uparrow\tilde{v}\downarrow\tilde{c}\uparrow\tilde{c}\downarrow;\tilde{v}\uparrow\tilde{v}\downarrow\tilde{c}\uparrow\tilde{c}\downarrow}=-v_{\tilde{c}\tilde{c}\tilde{c}\tilde{c}}+4v_{\tilde{v}\tilde{c}\tilde{c}\tilde{v}}-2v_{\tilde{v}\tilde{c}\tilde{v}\tilde{c}}-v_{\tilde{v}\tilde{v}\tilde{v}\tilde{v}},\\
 &\tilde{\Sigma}^{4p}_{\tilde{c}\uparrow\tilde{c}\downarrow\tilde{v}\uparrow\tilde{v}\downarrow;\tilde{c}\uparrow\tilde{c}\downarrow\tilde{v}\uparrow\tilde{v}\downarrow}=-\tilde{\Sigma}^{4p}_{\tilde{v}\uparrow\tilde{v}\downarrow\tilde{c}\uparrow\tilde{c}\downarrow;\tilde{v}\uparrow\tilde{v}\downarrow\tilde{c}\uparrow\tilde{c}\downarrow},
\end{align}
where the numerical values of $v_{\tilde{i}\tilde{j}\tilde{k}\tilde{l}}$ are reported in App.~\ref{App:eef_H}. 

Diagonalization of $\bar{H}^{\text{eff}}_4$ yields three resonant and three antiresonant excitation energies. The resonant energies are two singlets (one single excitation and one double excitation) and a triplet. They are reported in Table \ref{TableI} and compared to the exact energies. For completeness we also report in the table the excitation energies obtained using the standard (static) BSE which is built on a one-shot $GW$ calculation (BSE@GW). Note that the starting point of the one-shot $GW$ is the HF 1-GF and we use a static $W$ to calculate the $GW$ quasiparticles. The $(4,0)$-MCDE gives very good results for the (singlet and triplet) single excitations, with the triplet energy better described than within the BSE@GW method. We note that applying the Tamm-Dancoff approximation (TDA) to the head of $\bar{H}^{\text{eff}}_4$ improves even more the triplet energy for all the methods studied, whereas the singlet energy improves only in the case of MCDE@Exp (where the HF quasiparticle band gap in $L^{0,\text{4p}}$ is replaced by the experimental quasiparticle band gap).

As expected the  $(4,0)$-MCDE is able to reproduce also the double excitation, which is absent in the static BSE@GW approach.
Moreover, no spurious unphysical energies are obtained contrary to the dynamical BSW@GW method \cite{Rom09-2}.
However, the double excitation is overestimated with respect to the exact value by about $30\%$. 
One could improve results by dressing $L^{0,4\text p}$, i.e. uses an IP 1-GF beyond HF, and/or by going beyond the RPAx approximation for $\Sigma_4$.
We first explore the $L^{0,4\text p}$ dressing by using the $GW$ $L^{\text{0,4p}}_{4}$ instead of the HF $L^{\text{0,4p}}_{4}$. By simply substituting the HF quasiparticle band gap (32.52 eV) with the GW quasiparticle band gap (27.92 eV) in $L^{\text{0,4p}}_{4}$ greatly improves the agreement of the double excitation energy as well as the single singlet excitation energy with the exact result. We also note that by substituting the HF quasiparticle band gap with the experimental quasiparticle band gap of the Helium atom (24.50 eV) \cite{Brehm_67,Martin_87,Rom09-2} in $L^{\text{0,4p}}_{4}$ improves even more the double excitation energy. Instead, indroducing screening in $\Sigma_4$ has a negligible effect for this model ( results not reported in Table \ref{TableI}). 
These results point to the importance of dressing $L^{\text{0,4p}}_{4}$ in order to get accurate double excitations. 
\begin{table}[h]
	\caption{Triplet ($\omega^T_1$) and singlet single ($\omega^{S(S)}_1$) and double ($\omega^{S(D)}_1$) excitation energies of the He-like atom: exact $vs$ results obtained using the standard (static) BSE based on $GW$ (BSE@GW), the MCDE based on $L^{\text{HF}}_4$ (MCDE@HF), the MCDE based on  $L^{\text{GW,4p}}_4$ (MCDE@GW), and the MCDE based on $G^{\text{Exp,4p}}_4$, which is built using the experimantal quasiparticle band gap from Ref.~\onlinecite{Brehm_67,Martin_87,Rom09-2}. See text for details.}
	\begin{center}
			
			\begin{tabular}{c||ccc}
				\small
				Method & $\omega^T_1$ & $\omega^{S(S)}_2$ &$\omega^{S(D)}_3$ \\ \hline
			Exact & 19.22 & 23.77 & 58.02 \\
			BSE@GW & 14.79 & 23.58 & -  \\
			MCDE@HF & 18.74 & 24.05 & 75.73  \\
			MCDE@GW & 18.74 & 23.77& 66.80  \\
            MCDE@Exp & 18.74 & 23.48& 60.26  \\
            BSE@GW@TDA & 15.39 & 23.97 & -  \\
            MCDE@HF@TDA & 19.02  & 24.26 & 75.73  \\
			MCDE@GW@TDA & 19.02 & 23.98& 66.81  \\
            MCDE@Exp@TDA & 19.02  & 23.69& 60.27  \\
			\end{tabular}
	\end{center}	
	\label{TableI}
\end{table}

\section{Conclusions\label{sec:conclusions}}
In this work we generalized the MCDE, recently derived to couple the one-body and the three-body Green's function, to the coupling of even-order Green's functions and to the coupling of odd-order Green's functions. As an example we focused on the coupling between the 2- and the 4-body GFs. We studied the electron-hole channel of the 2-GF, but along a similar line the MCDE can be derived for the particle-particle channel. Using the same recipe as for the 1-GF and 3-GF coupling, we showed how to systematically build a static approximation for the 4-body self-energy as well as how to map the MCDE to an eigenvalue problem. As an illutsration we applied our method to the study of neutral excitations in a simple Helium-like model with two levels. Our approach can overall very well reproduce the two singlet (one single and one double exitation) and the triplet energies, unlike the standard BSE@GW approach, which, in its static approximation misses the double excitation. Several research lines based on the concepts developed in this work can be explored in the future. For example, one could analyze the two-body self-energy which results from the solution of the MCDE and use its static approximation to improve single excitations. Another interesting perspective is to derive the MCDE for the particle-particle channel of the 2-GF. Recently the particle-particle BSE beyond RPA has been derived using anomalous propagators;\cite{Marie_PRB2024,Marie_arxiv2024} it would be interesting to compare the two approaches. More in general our MCDE could be also employed to study explicit n-particle excitations, such as trions,\cite{Deilmann_2016,Torche_2019}, RIXS,\cite{Vorwerk_2022} photoemission from an excitonic state \cite{Perfetto_2016} etc.

\textit{Acknowledgment:}
We thank the French “Agence Nationale de la Recherche (ANR)” for financial support (Grant Agreements No. ANR-19-CE30-0011 and No. ANR-22-CE30-0027).
\appendix
\section{The independent-particle 4-GF\label{App:NI_G4}}
Let us analyze the IP 4-GF given in Eq.~\eqref{G04wick:eq}. 
Each term is composed of a product of four IP 1-GF. 
Using the time diferences given in Eq.~\eqref{timesG4:eq} we can distinguish 3 cases with each case having a different number of noninteracting 1-GFs in the product that depend on the time difference $\tau = t_1 - t_3$. We can have either 0, 2, our 4 of those noninteracting 1-GFs. The other noninteracting 1-GFs in the products are time independent.
Les us discuss an example of each of these 3 cases.

1) An example of a contribution to $G_4^0$ composed of a product of 4 noninteracting 1-GFs all of which depend on $\tau$ is given by
\begin{align}
    &G_1^0(1,4')G_1^0(2,3') G_1^0(3,2') G_1^0(4,1')= \nonumber\\
    &G_1^0(x_1,x_{4'};\tau)G_1^0(x_2,x_{3'};\tau) G_1^0(x_3,x_{2'};-\tau) G_1^0(x_4,x_{1'};-\tau).
\end{align}
It describes the propagation of 4 noninteracting particles.
Its spectral representation is given by
\begin{align}\label{GGGGdinamical:eq}
   & i[G_1^0(x_1,x_{4'})G_1^0(x_2,x_{3'}) G_1^0(x_3,x_{2'}) G_1^0(x_4,x_{1'})](\omega)=\nonumber \\
    &\sum_{\substack{v,v' \\ c,c'}}\!\!\frac{\phi_c(x_1)\phi_c^*(x_{4'})\phi_{c'}(x_2)\phi_{c'}^*(x_{3'})\phi_{v'}(x_3)\phi_{v'}^*(x_{2'})\phi_{v}(x_4)\phi_{v}^*(x_{1'})}{\omega - (\epsilon^0_c-\epsilon^0_v) - (\epsilon^0_{c'}-\epsilon^0_{v'})+i\eta}\nonumber\\
    &-\sum_{\substack{v,v' \\ c,c'}}\!\!\dfrac{\phi_v(x_1)\phi_v^*(x_{4'})\phi_{v'}(x_2)\phi_{v'}^*(x_{3'})\phi_{c'}(x_3)\phi_{c'}^*(x_{2'})\phi_{c}(x_4)\phi_{c}^*(x_{1'})}{\omega + (\epsilon^0_c-\epsilon^0_v) + (\epsilon^0_{c'}-\epsilon^0_{v'})-i\eta}.
\end{align}
From this expression, we clearly see that the poles of this term correspond to a sum of two eigenenergy differences of a conduction and a valence state. This shows that (approximate) double neutral excitations are already present at the independent-particle level. 

2) An example of a contribution to $G_4^0$ composed of a product of 4 noninteracting 1-GFs of which 2 depend on $\tau$ is given by
\begin{align}
        & -G_1^0(1,1')G_1^0(2,3') G_1^0(3,2') G_1^0(4,4')= \nonumber \\ 
        &-G_1^0(x_1,x_{1'};0^+)G_1^0(x_2,x_{3'};\tau) G_1^0(x_3,x_{2'};-\tau) G_1^0(x_4,x_{4'};0^+).
\end{align}
It describes the propagation of 2 noninteracting particles.
The two time dependent $G^0_1$ define the poles of the 4-GF in accordance with the $eh$ channel of the 2-GF, i.e. $G^{\text{eh}}_2(1,2,x_{1'}t_1,x_{2'}t_2)$. Its spectral representation reads
\begin{align}
        & -G_1^0(x_1,x_{1'};0^+)G_1^0(x_2,x_{3'};\tau) G_1^0(x_3,x_{2'};-\tau) G_1^0(x_4,x_{4'};0^+) \nonumber \\ 
        &=-iG_1^0(x_1,x_{1'};0^+)G_1^0(x_4,x_{4'};0^+)\times \nonumber\\ &\Bigg[\sum_{v',c}\frac{\phi_c(x_2)\phi^*_c(x_{3'})\phi_{v'}(x_3)\phi^*_{v'}(x_{2'})}{\omega-(\epsilon_c-\epsilon_{v'})-i\eta}\nonumber\\
        &-\sum_{v,c'}\frac{\phi_{c'}(x_3)\phi^*_{c'}(x_{2'})\phi_{v}(x_2)\phi^*_{v}(x_{3'})}{\omega+(\epsilon_{c'}-\epsilon_v)+i\eta}\Bigg].
\end{align}

3) Finally, an example of a contribution to $G_4^0$ composed of a product of 4 noninteracting 1-GFs of which none depend on $\tau$ is given by
\begin{align}
    &G_1^0(1,1')G_1^0(2,2')G_1^0(3,3') G_1^0(4,4')=\nonumber \\
    &G_1^0(x_1,x_{1'};0^+)G_1^0(x_2,x_{2'};0^+) G_1^0(x_3,x_{3'};0^+) G_1^0(x_4,x_{4'};0^+),
\end{align}
which does not describe any propagation.
Such terms do not have poles and, therefore, they can be excluded. 
\section{$(4,0)$-MCDE in real space}\label{sec:realspace}
Following the change of basis in Eq.~\eqref{G4changebasis:eq}, the multichannel self-energy in Eqs.~\eqref{selfbody4:eq}-\eqref{selftildecoupling4:eq} becomes 
\begin{align}
 \Sigma^{2p}(x_{2},x_{3},x_{2'},x_{3'}) &=\delta(x_{2},x_{2'})\delta(x_{3},x_{3'})v(\mathbf{r}_{2},\mathbf{r}_{3})\nonumber\\
 &-\delta(x_{3},x_{2'})\delta(x_{2},x_{3'})v(\mathbf{r}_{2},\mathbf{r}_{3}),
\end{align}
\begin{widetext}\begin{align}\label{self4_couplingreal:eq}
    \Sigma^\text{2p/4p}(x_2,x_3,x_4,x_{2'},x_{3'},x_{4'})&\!=\!
    \delta(x_{2},x_{3'})[\delta(x_{2'}\!,x_{3})\delta(x_{4'}\!,x_{4})\!-\!\delta(x_{2'}\!,x_{4})\delta(x_{4'}\!,x_{3})]v(\mathbf{r}_{2'}\!,\!\mathbf{r}_{4'}\!)
    \nonumber\\
    &\!+\!\delta(x_{2'}\!,x_{3})[\delta(x_{2},x_{4'})\delta(x_{4},x_{3'})\!-\!\delta(x_{2},x_{3'})\delta(x_{4},x_{4'})]v(\mathbf{r}_{2},\mathbf{r}_{4})
    \nonumber\\
    &\!+\!\delta(x_{2'}\!,x_{4})[\delta(x_{2},x_{3'})\delta(x_{3},x_{4'})\!-\!\delta(x_{2},x_{4'})\delta(x_{3},x_{3'})]v(\mathbf{r}_{2},\mathbf{r}_{3})
    \nonumber\\
    &\!+\!\delta(x_{2},x_{4'})[\delta(x_{2'}\!,x_{4})\delta(x_{3'}\!,x_{3})\!-\!\delta(x_{2'}\!,x_{3})\delta(x_{3'}\!,x_{4})]v(\mathbf{r}_{2'}\!,\!\mathbf{r}_{3'}\!),
\end{align}\end{widetext}

\begin{widetext}\begin{align}\label{self4_couplingtildereal:eq}
    \Sigma^\text{4p/2p}(x_1,x_2,x_3,x_{1'},x_{2'}x_{3'})&\!=\!
    \delta(x_{2},x_{3'})[\delta(x_{1},x_{1'})\delta(x_{3},x_{2'})\!-\!\delta(x_{1},x_{2'})\delta(x_{3},x_{1'})]v(\mathbf{r}_{1},\mathbf{r}_{3})
    \nonumber\\
    &\!+\!\delta(x_{2'},x_{3})[\delta(x_{1'}\!,x_{2})\delta(x_{3'}\!,x_{1})\!-\!\delta(x_{1'}\!,x_{1})\delta(x_{3'}\!,x_{2})]v(\mathbf{r}_{1'},\mathbf{r}_{3'})
    \nonumber\\
    &\!+\!\delta(x_{1'},x_{3})[\delta(x_{2'}\!,x_{1})\delta(x_{3'}\!,x_{2})\!-\!\delta(x_{2'}\!,x_{2})\delta(x_{3'}\!,x_{1})]v(\mathbf{r}_{2'},\mathbf{r}_{3'})
    \nonumber\\
    &\!+\!\delta(x_{1},x_{3'})[\delta(x_{2},x_{2'})\delta(x_{3},x_{1'})\!-\!\delta(x_{2},x_{1'})\delta(x_{3},x_{2'})]v(\mathbf{r}_{2},\mathbf{r}_{3}),
\end{align}\end{widetext}

\begin{widetext}
\begin{align}\label{self4_bodyreal:eq}
    \Sigma^{\text{4p}}&(x_1,x_2,x_3,x_4,x_{1'},x_{2'},x_{3'},x_{4'})=\nonumber\\
    &= \delta(x_{1},x_{4'})\delta(x_{2},x_{3'})[\delta(x_{2'},x_{4})\delta(x_{1'},x_{3})-\delta(x_{2'},x_{3})\delta(x_{1'},x_{4})]v(\mathbf{r}_{2'},\mathbf{r}_{1'}) 
    \nonumber\\
    &+ \delta(x_{2'},x_{3})\delta(x_{1'},x_{4})[\delta(x_{1},x_{3'})\delta(x_{2},x_{4'})-\delta(x_{1},x_{4'})\delta(x_{2},x_{3'})]v(\mathbf{r}_{1},\mathbf{r}_{2})  
   \nonumber\\
    &+ \delta(x_{2'},x_{4})\delta(x_{1'},x_{3})[\delta(x_{1},x_{4'})\delta(x_{2},x_{3'})-\delta(x_{1},x_{3'})\delta(x_{2},x_{4'})]v(\mathbf{r}_{1},\mathbf{r}_{2})  
   \nonumber\\
    &+ \delta(x_{1},x_{3'})\delta(x_{2},x_{4'})[\delta(x_{2'},x_{3})\delta(x_{1'},x_{4})-\delta(x_{2'},x_{4})\delta(x_{1'},x_{3})]v(\mathbf{r}_{1'},\mathbf{r}_{2'})  
   \nonumber\\
    &+ \delta(x_{1},x_{4'})\delta(x_{1'},x_{4})[\delta(x_{2},x_{3'})\delta(x_{3},x_{2'})-\delta(x_{2},x_{2'})\delta(x_{3},x_{3'})]v(\mathbf{r}_{2},\mathbf{r}_{3})  
   \nonumber\\
    &+ \delta(x_{1},x_{4'})\delta(x_{2'},x_{3})[\delta(x_{2},x_{3'})\delta(x_{4},x_{1'})-\delta(x_{2},x_{1'})\delta(x_{4},x_{3'})]v(\mathbf{r}_{2},\mathbf{r}_{4})  
   \nonumber\\
    &+ \delta(x_{2},x_{3'})\delta(x_{1'},x_{4})[\delta(x_{1},x_{4'})\delta(x_{3},x_{2'})-\delta(x_{1},x_{2'})\delta(x_{3},x_{4'})]v(\mathbf{r}_{1},\mathbf{r}_{3})  
   \nonumber\\
    &+ \delta(x_{2},x_{3'})\delta(x_{2'},x_{3})[\delta(x_{1},x_{4'})\delta(x_{4},x_{1'})-\delta(x_{1},x_{1'})\delta(x_{4},x_{4'})]v(\mathbf{r}_{1},\mathbf{r}_{4})  
   \nonumber\\
    &- \delta(x_{1},x_{4'})\delta(x_{2'},x_{4})[\delta(x_{2},x_{3'})\delta(x_{3},x_{1'})-\delta(x_{2},x_{1'})\delta(x_{3},x_{3'})]v(\mathbf{r}_{2},\mathbf{r}_{3})  
   \nonumber\\
    &- \delta(x_{1},x_{4'})\delta(x_{1'},x_{3})[\delta(x_{2},x_{3'})\delta(x_{4},x_{2'})-\delta(x_{2},x_{2'})\delta(x_{4},x_{3'})]v(\mathbf{r}_{2},\mathbf{r}_{4})  
   \nonumber\\
    &- \delta(x_{2},x_{3'})\delta(x_{2'},x_{4})[\delta(x_{1},x_{4'})\delta(x_{3},x_{1'})-\delta(x_{1},x_{1'})\delta(x_{3},x_{4'})]v(\mathbf{r}_{1},\mathbf{r}_{3})  
   \nonumber\\
    &- \delta(x_{2},x_{3'})\delta(x_{1'},x_{3})[\delta(x_{1},x_{4'})\delta(x_{4},x_{2'})-\delta(x_{1},x_{2'})\delta(x_{4},x_{4'})]v(\mathbf{r}_{1},\mathbf{r}_{4})  
   \nonumber\\
    &- \delta(x_{1},x_{3'})\delta(x_{1'},x_{4})[\delta(x_{2},x_{4'})\delta(x_{3},x_{2'})-\delta(x_{2},x_{2'})\delta(x_{3},x_{4'})]v(\mathbf{r}_{2},\mathbf{r}_{3})  
   \nonumber\\
    &- \delta(x_{1},x_{3'})\delta(x_{2'},x_{3})[\delta(x_{2},x_{4'})\delta(x_{4},x_{1'})-\delta(x_{2},x_{1'})\delta(x_{4},x_{4'})]v(\mathbf{r}_{2},\mathbf{r}_{4})  
    \nonumber\\
    &- \delta(x_{2},x_{4'})\delta(x_{1'},x_{4})[\delta(x_{1},x_{3'})\delta(x_{3},x_{2'})-\delta(x_{1},x_{2'})\delta(x_{3},x_{3'})]v(\mathbf{r}_{1},\mathbf{r}_{3})  
   \nonumber\\
    &- \delta(x_{2},x_{4'})\delta(x_{2'},x_{3})[\delta(x_{1},x_{3'})\delta(x_{4},x_{1'})-\delta(x_{1},x_{1'})\delta(x_{4},x_{3'})]v(\mathbf{r}_{1},\mathbf{r}_{4})  
   \nonumber\\
    &+ \delta(x_{1},x_{3'})\delta(x_{2'},x_{4})[\delta(x_{2},x_{4'})\delta(x_{3},x_{1'})-\delta(x_{2},x_{1'})\delta(x_{3},x_{4'})]v(\mathbf{r}_{2},\mathbf{r}_{3})  
   \nonumber\\
    &+ \delta(x_{1},x_{3'})\delta(x_{1'},x_{3})[\delta(x_{2},x_{4'})\delta(x_{4},x_{2'})-\delta(x_{2},x_{2'})\delta(x_{4},x_{4'})]v(\mathbf{r}_{2},\mathbf{r}_{4})  
   \nonumber\\
    &+ \delta(x_{2},x_{4'})\delta(x_{2'},x_{4})[\delta(x_{1},x_{3'})\delta(x_{3},x_{1'})-\delta(x_{1},x_{1'})\delta(x_{3},x_{3'})]v(\mathbf{r}_{1},\mathbf{r}_{3})  
   \nonumber\\
    &+ \delta(x_{2},x_{4'})\delta(x_{1'},x_{3})[\delta(x_{1},x_{3'})\delta(x_{4},x_{2'})-\delta(x_{1},x_{2'})\delta(x_{4},x_{3'})]v(\mathbf{r}_{1},\mathbf{r}_{4})  .
\end{align}
\end{widetext}

\section{Calculation of the 4-body self-energy elements\label{App:eef_H}}
In this section we report the expression of the elements $v_{\tilde{i}\tilde{j}\tilde{k}\tilde{o}}$ which builds $\Sigma_4$ in terms of the bonding/antibonding basis ($\phi_v/\phi_c$) set which diagolanizes the non interacting Hamiltonian. They read:
\begin{widetext}
\begin{align}
	\label{eq:V_HF_start}
	v_{\tilde{v}\tilde{v}\tilde{v}\tilde{v}}&=\int\text{d}x\text{d}y
	\left[\alpha \phi_v + \beta \phi_c\right]^\ast(x)
	\left[\alpha \phi_v + \beta \phi_c\right]^\ast(y)
	v(x,y)
	\left[\alpha \phi_v + \beta \phi_c\right](y)
	\left[\alpha \phi_v + \beta \phi_c\right](x)
	\\
	&=\alpha^4 v_{vvvv}
	+ 4\beta\alpha^3v_{cvvv}
	+ 4 \beta^2\alpha^2 v_{cvcv}
	+2\beta^2 \alpha^2 v_{cvvc}
	+4 \beta^3\alpha v_{cccv}
	+\beta^4 v_{cccc}
	\\
	v_{\tilde{c}\tilde{c}\tilde{c}\tilde{c}}&=
	\int\text{d}x\text{d}y
	\left[\beta \phi_v - \alpha \phi_c\right]^\ast(x)
	\left[\beta \phi_v - \alpha \phi_c\right]^\ast(y)
	v(x,y)
	\left[\beta \phi_v - \alpha \phi_c\right](y)
	\left[\beta \phi_v - \alpha\phi_c\right](x)
	\\
	&=\beta^4 v_{vvvv}
	- 4\alpha\beta^3v_{cvvv}
	+ 4 \alpha^2\beta^2 v_{cvcv}
	+2\alpha^2 \beta^2 v_{cvvc}
	-4 \alpha^3\beta v_{cccv}
	+\alpha^4 v_{cccc}
	\\
	v_{\tilde{v}\tilde{v}\tilde{c}\tilde{c}}&=
	\int\text{d}x\text{d}y
	\left[\alpha \phi_v + \beta \phi_c\right]^\ast(x)
	\left[\alpha \phi_v + \beta \phi_c\right]^\ast(y)
	v(x,y)
	\left[\beta \phi_v -\alpha \phi_c\right](y)
	\left[\beta \phi_v -\alpha \phi_c\right](x)\nonumber
	\\
	&=\beta ^2 \alpha ^2 v_{cccc}+ \left(2 \alpha^3  \beta
	-2 \beta ^3\alpha \right)v_{cccv}-2 \alpha^2  \beta^2 v_{cvvc}+ \left(\alpha ^4 -2
	\alpha^2  \beta^2+\beta ^4\right)v_{vvcc}+ \left(-2 \alpha ^3 \beta+2 \alpha 
	  \beta ^3\right)v_{vvvc}\nonumber\\&+\alpha ^2 \beta ^2 v_{vvvv}
	\\
	v_{\tilde{v}\tilde{v}\tilde{v}\tilde{c}}&=
	\int\text{d}x\text{d}y
	\left[\alpha \phi_v + \beta \phi_c\right]^\ast(x)
	\left[\alpha \phi_v + \beta \phi_c\right]^\ast(y)
	v(x,y)
	\left[\alpha \phi_v + \beta \phi_c\right](y)
	\left[\beta \phi_v -\alpha \phi_c\right](x)\nonumber
	\\
	&=-\alpha \beta ^3  v_{cccc}+ \left(-3 \alpha^2  \beta
	^2 +\beta ^4 \right)v_{cccv}+ \left(-\alpha ^3
	\beta +\alpha  \beta ^3 \right)v_{cvvc}+
	\left(-2 \alpha ^3 \beta +2 \alpha  \beta ^3 	\right)v_{vvcc}+ \left(3 \alpha ^2 \beta^2  -\alpha ^4
\right)v_{vvvc}\nonumber\\
  &+\alpha ^3 \beta  v_{vvvv}
	\\
	v_{\tilde{v}\tilde{c}\tilde{c}\tilde{v}}&=
	\int\text{d}x\text{d}y
	\left[\alpha \phi_v + \beta \phi_c\right]^\ast(x)
	\left[\beta \phi_v - \alpha \phi_c\right]^\ast(y)
	v(x,y)
	\left[\beta \phi_v -\alpha \phi_c\right](y)
	\left[ \alpha\phi_v + \beta\phi_c\right](x)\nonumber
	\\
	&=
	\beta ^2 \alpha ^2 v_{cccc}+ \left(2 \alpha ^3 \beta
	-2 \beta ^3 \alpha \right)v_{cccv}+
	\left(\alpha ^24+\beta ^4 \right)v_{cvvc}-4 \alpha^2  \beta^2 
	v_{vvcc}+ \left(-2 \alpha ^3
	\beta +2 \alpha  \beta^3\right)v_{vvvc}\nonumber\\
&+\alpha ^2 \beta ^2
	v_{vvvv}
	\\
	v_{\tilde{c}\tilde{c}\tilde{c}\tilde{v}}&=
	\int\text{d}x\text{d}y
	\left[\beta \phi_v -\alpha \phi_c\right]^\ast(x)
	\left[\beta \phi_v -\alpha\phi_c\right]^\ast(y)
	v(x,y)
	\left[\beta \phi_v -\alpha \phi_c\right](y)
	\left[\alpha \phi_v + \beta \phi_c\right](x)\nonumber
	\\
	&=
	\label{eq:v_HF_end}
	-\beta  \alpha ^3 v_{cccc}+ \left(-\alpha^4  +3 \beta^2\alpha ^2\right)v_{cccv}+
	\left(\alpha^3 \beta-\beta^3 \alpha \right)v_{cvvc}+\left(2 \alpha^3  \beta-2 \alpha\beta^3 \right)  v_{vvcc}
	+ \left(\beta^4  -3 \alpha^2 \beta ^2 \right)v_{vvvc}\nonumber\\&+\alpha \beta ^3 v_{vvvv}
\end{align}   
\end{widetext}
%

\end{document}